\documentclass[twocolumn,showpacs,preprintnumbers,amsmath,amssymb]{revtex4}

\usepackage{graphicx}
\usepackage{dcolumn}
\usepackage{bm}

\begin{document}

\preprint{APS/123-QED}

\title{An accurate scheme to calculate the interatomic \\
  Dzyaloshinskii-Moriya interaction parameters}

\author{}
\author{S.~Mankovsky and H.~Ebert}
\affiliation{%
$^1$Department of Chemistry/Phys. Chemistry, LMU Munich,
Butenandtstrasse 11, D-81377 Munich, Germany 
}%

\date{\today}

\begin{abstract}
An new and accurate scheme to calculate the
interatomic Dzyaloshinskii-Moriya interaction (DMI)
parameters is presented, which is based on the
fully relativistic
 Korringa-Kohn-Rostoker Green function 
(KKR-GF) technique. 
Corresponding numerical results are compared with
those obtained using other schemes reported in the literature. 
 The differences found
 can be attributed primarily to the
different reference states used in the various approaches. 
In addition 
 an expression for the DMI parameters formulated for a micromagnetic model
 Hamiltonian is presented that provides a connection
to the DMI parameters calculated for atomistic Hamiltonians.  
This formulation also allows the 
discussion of the DMI in terms of specific features of the electronic
band structure.
 
\end{abstract}

\maketitle

\section{ Introduction \label{IN}}

Recent investigations on the influence of spin-orbit coupling (SOC)  on the
magnetic and transport properties of solids open the
way  for an efficient tuning of these properties 
concerning their application in various types of electronic devices
based on new technologies. 
This concerns in particular new phenomena 
associated with a chiral magnetic texture, as for example skyrmions
\cite{BH94,RBP06} with the related topological and anomalous Hall effects
\cite{SRB+12,NPB+09} or chiral magnetic soliton lattices \cite{TKT+12} with
corresponding magneto-resistance phenomena \cite{TKN+13}. 
In fact, it is the SOC induced anisotropic
Dzyaloshinskii-Moriya exchange
 interaction (DMI) that is
 responsible for the creation of
such non-trivial magnetic textures in  non-centrosymmetric systems. 
As in many other fields of material science
the search for materials with corresponding favorable 
 isotropic and anisotropic exchange coupling parameters
can be substantially supported by
   first-principles 
investigations.
 Actually, several schemes   to calculate 
the DMI parameters of a solid have been suggested
with their explicit 
formulation depending on  the underlying
electronic structure method. 

A rather flexible approach  to calculate
 the exchange coupling
parameters is based on the  Korringa-Kohn-Rostoker Green function 
(KKR-GF) technique in its multiple
scattering formulation. In fact two such schemes, scheme I \cite{USPW03}
and  scheme II \cite{EM09a}, based on the  magnetic force 
theorem have been reported 
that can be seen as a relativistic generalization of the so-called Lichtenstein formula \cite{LKAG87}
giving this way not only access to the isotropic exchange coupling
parameter but to the exchange coupling tensor
$\underline{J}_{ij}$. 
The components of the corresponding DMI 
vector $\vec{D}$  are determined by 
the anti-symmetric part of this tensor by a linear combination of the off-diagonal elements of
$\underline{J}_{ij}$ (see below).  

The mentioned KKR-GF based approaches allow a
 direct
calculation of the exchange coupling tensor
$\underline{J}_{ij}$  in real space with its elements given by
the
variation of single particle energy caused by an infinitesimal rotation of
two magnetic moments $\vec m_i$ and  $\vec m_j$ 
 on the atomic sites $i$ and $j$, respectively.  
An alternative approach suggested in the literature is focused
 on the relevant parameters of  micromagnetic models,
 i.e.\ the exchange
stiffness $A$ and the 
DMI vector $\vec{D}$ \cite{HBB08,HBB09},
that are obtained  by fitting the 
spin-wave dispersion curve $\omega(\vec {q})$ 
 calculated from  first-principles using expressions 
based  on these model parameters. 
A rather elegant way for the calculation of the micromagnetic DMI vector  
was suggested recently by Freimuth et al.\ \cite{FBM14,KNA15} 
 exploiting a  property of the spin wave spectra
of  non-centrosymmetric systems. 
While the corresponding spectra of
centrosymmetric systems have a parabolic like dispersion around their
minimum at the $\Gamma$ point ($\vec{q} = 0$),  the minimum
moves away from the $\Gamma$ point for non-centrosymmetric systems due to  the DMI with the dispersion
at $\vec{q} = 0$ becoming non-zero. 
This feature can be used to map the
calculated electronic energy connected with  a spin spiral configuration
in the system to  the microscopic Heisenberg model Hamiltonian,  
giving this way access to the DMI parameters \cite{FBM14}. 

Starting from the scheme of Freimuth et al.\ \cite{FBM14}, a new scheme
(scheme III) to calculate the DMI parameters for any pair of atoms  is suggested here
that is  based on the KKR-GF method. As it will  be demonstrated, the
advantage of this approach 
is that  it allows for any orientation of the magnetization the simultaneous
 calculation of the two components of the DMI vector perpendicular to the magnetization. 
These are the only components of $\vec{D}$ which can be
determined as the magnetic moments
 have only two possible degrees of freedom to 
deviate from the magnetization direction.
 This implies also that the 
component of the DMI vector along the magnetization direction is not defined.
The approach to be presented allows also a KKR-GF
based formulation for
the micromagnetic DMI,  permitting  a comparison 
of the DMI vectors formulated within
two different approaches, but calculated using the same
electronic structure method.

\section{Theoretical formulation \label{THEO}}

\subsection{Hamiltonians}


In order to derive an expression  for the interatomic DMI vector $\vec{D}_{ij}$
we start from a fully relativistic description of the electronic structure of a magnetic solid on the basis of the  Dirac Hamiltonian
 \begin{eqnarray}
 {\cal H}_{\rm D}   & =&
 - i c \vec{\alpha} \cdot \vec{\nabla}  
+ \frac{1}{2} \, c^{2} (\beta - 1) 
\nonumber \\ 
&&\qquad 
+ \bar V(\vec {r}) + \beta \, \vec{\sigma} \cdot  \vec { B}(\vec {r})  
 + e \vec{\alpha} \cdot \vec {A}(\vec {r})
\;,
\label{Eq:Dirac}
\end{eqnarray}
that is setup in the framework of  relativistic
 spin density functional theory \cite{ED11}. 
Accordingly, 
$ {\alpha}_i $ and $ \beta $ are the standard Dirac matrices \cite{Ros61}  while
 $\bar V(\vec {r}) $ and 
$   \vec { B}(\vec {r}) $ are the spin independent and dependent parts of the electronic potential.

When adopting a simplified atomistic microscopic model approach, an
expression for the
exchange coupling tensor 
$\underline{J}_{ij}$ of the generalized Heisenberg Hamiltonian
 \begin{eqnarray}
 {\cal H}_{\rm H}   & =&
\sum_{i j}  \hat m_{i} \,  \underline{J}_{ij} \,  \hat m_{j} 
\label{Eq:Heisenberg}
\end{eqnarray}
can be derived by
  mapping the magnetic energy obtained within
electronic structure calculations based on 
Eq.\ (\ref{Eq:Dirac}) to the energy corresponding to this model Hamiltonian. 
 Here, we focus on the contribution 
\begin{eqnarray}
  {\cal H}_{\rm DM}  &=& \sum_{ij} \vec{D}_{ij} \cdot (\hat{m}_i \times
  \hat{m}_j) 
\label{Eq:Heisenberg_DMI}
\end{eqnarray}
connected with the DMI between the magnetic moments $\vec m_i$ and
$\vec m_j$, that is determined by the DMI vector $\vec{D}_{ij} =
(D^x_{ij}, D^y_{ij}, D^z_{ij})$ with 
\begin{eqnarray}
D^{\gamma}_{ij} &=&
\epsilon^{\alpha\beta\gamma}  \frac{J^{\alpha\beta}_{ij} -
  J^{\beta\alpha}_{ij}}{2} \; .
\end{eqnarray}
 
In the micromagnetic approach, the magnetic state is characterized by the
free energy density (omitting the magnetic anisotropy term)
\begin{eqnarray}
  F(\vec{r}) &=& \sum_{\mu,\nu} A^{\nu}\left(\frac{\partial m_\mu}{\partial r_{\nu}}\right)^2 + \sum_{\mu} \vec{D}^\mu \cdot
  \bigg(\hat{m} \times  \frac{\partial \hat{m}}{\partial r_\mu} \bigg)\;, 
\label{Eq:Micromanetic}
\end{eqnarray}
that is a functional of the continuous magnetization field $\vec{m}\equiv
\vec{m}(\vec{r})$. The second term in Eq.\ (\ref{Eq:Micromanetic}) describes the
micromagnetic energy density due to the DMI.

\subsection{DMI vector within the atomistic approach \label{THEO_II}}

To determine the DMI vector $\vec{D}$, we exploit the fact that the DMI
determine the slope of the  
 dispersion curve $\omega(\vec {q})$ of spin waves at the $\Gamma$ point.
Accordingly, 
 in order to find the $y$ component of the  DMI vector, 
we consider a  spin spiral
 with the  spin moments $\vec m_i$  rotating
 within the $x-z$ plane and with the wave vector
 $\vec{q}$ perpendicular to this plane, represented by the expression 
\begin{equation}
  \hat{m}_i =  \Big(\sin(\vec{q}\cdot\vec{R}_i),0, \cos(\vec{q}\cdot\vec{R}_i)\Big) 
\; ,
\label{spin-spiral_1}
\end {equation}
with  $\vec{q} = (0,q,0)$.

According to the Hamiltonian in Eq. (\ref{Eq:Heisenberg}), the
contribution to the energy of a spin spiral state due to the DMI
with respect to a collinear ferromagnetic reference state is 
given by:
%
\begin{eqnarray}
  E^{(1)}_{\rm DM} &=&  \sum_{ij} \vec{D}_{ij} \cdot (\hat{m}_i \times \hat{m}_j)
  \nonumber  \\
  & = & \sum_{ij} D^y_{ij} \,(m^z_i m^x_j  - m^x_i m^z_j)
  \nonumber  \\
  &=&  \sum_{ij} D^y_{ij}\, \sin(\vec{q}\cdot(\vec{R}_j - \vec{R}_i))  
\; .
\label{Heisenberg}
\end{eqnarray}
%
Although a spin spiral structure gives  also rise to
 other contributions
to the energy change, e.g.\ due to the isotropic exchange interaction, their
derivative with respect to the $\vec{q}$ vector vanishes at $\vec{q} =
0$. Accordingly, one has:
%
\begin{equation}
 \frac{\partial E^{(1)}}{\partial q_\alpha} \bigg\vert _{q \to 0}  =
 \frac{\partial E_{\rm DM}^{(1)}}{\partial q_\alpha} \bigg\vert _{q \to 0} \; .
\label{Limit_q0}
\end {equation}
%
Thus, the slope of the energy dispersion  of
a spin spiral described by  Eq.\ (\ref{spin-spiral_1})
is given  at $\vec{q} = 0$ by 
%
\begin{eqnarray}
  \lim_{q \to 0} \frac{\partial E_{\rm DM}^{(1)}}{\partial q_y}   
  &=& \lim_{q \to 0} \frac{\partial }{\partial q_y}   \sum_{ij} D^y_{ij} 
 \, \sin(\vec{q}\cdot(\vec{ R}_j - \vec{R}_i))  \nonumber \\  
  &=&  \sum_{ij} D^y_{ij} \, (\vec{R}_j - \vec{R}_i)_y    \;.
\label{Slope}
\end{eqnarray}

To map  the free energy  $F$ determined within the microscopic
representation onto the Heisenberg Hamiltonian in Eq.\
(\ref{Eq:Heisenberg}), one can start from the relationship between the
free energy operator ${\cal F}$ and the grand-canonical energy given in
operator form ${\cal K = H - \mu N }$:
%
\begin{eqnarray}
\frac{\partial }{\partial \beta}(\beta {\cal F})  
  &=&  {\cal  K} \nonumber \\
  &=&  {\cal H - \mu N }  \;,
\end{eqnarray}
with $\mu$ the chemical potential and 
    $  \beta = (kT)^{-1}$. This leads to the
following  expression for the  variation
of the corresponding  single-particle energy density
$K^{(1)}(\vec{r})$ associated with the 
spin-spiral structure in terms of the electronic Green function 
%
\begin{eqnarray}
K^{(1)}(\vec{r}) &=& -\frac{1}{\pi} \mbox{Im}\,\mbox{Tr}  \int^\mu dE\, [({\cal H} - \mu)\,
  G(\vec{r},\vec{r},E) \nonumber \\
&& \qquad \qquad \qquad  - ({\cal H}_0 - \mu) \,G_0(\vec{r},\vec{r},E) ] 
\label{K1_dens_new1} \nonumber  \\
&=&  -\frac{1}{\pi} \, \mbox{Im}\, \mbox{Tr} \int^\mu dE\, (E - \mu)\, \Delta
G(\vec{r},\vec{r},E) \;,
\label{K1_dens_new2}
\end{eqnarray}
%
where we restricted to the case $T = 0$ and used the index $0$ to indicate the collinear 
ferromagnetic reference  state.
Obviously, 
$\Delta G(\vec{r},\vec{r}\,',E)$ represents the change of  the Green function due to the
formation of a  spin-spiral structure described by Eq.\ (\ref{spin-spiral_1}). 
The corresponding perturbation $\Delta V$  giving rise to $\Delta G(\vec{r},\vec{r}\,',E)$ 
 is given by the change of the local effective
exchange-correlation field
$\vec{B}_{xc}(\vec{r})$ 
due to a rotation of the magnetic moments  $\vec{m}_i$
on sites $i$ away from the collinear reference direction 
$\hat{m}_0 = \hat{z}$.
According to  Eq.\ (\ref{Eq:Dirac})
one can write for the specific wave vector $\vec{q}$
\begin{eqnarray}
{\Delta V_{\vec{q}}(\vec{r})} &=&  
 \sum_i \delta V_{\vec{q}}(\vec{r} - \vec{R}_i)  \nonumber \\
&=& \beta \sum_i [\vec{\sigma}\cdot   \hat{m}_i - \sigma_z)]\,  
B_{xc}(\vec{r} - \vec{R}_i)  \nonumber  \\
&=& \beta \sum_i 
 [\sin (\vec{q} \cdot \vec{R}_i)\,\sigma_x  \nonumber \\
& & + (\cos (\vec{q} \cdot \vec{R}_i) - 1)\,\sigma_z]\, 
B_{xc}(\vec{r} - \vec{R}_i)   
\; .
\label{Eq:perturb}
\end{eqnarray}
The magnitude of the perturbation potential
given by Eq.\ (\ref{Eq:perturb}) is
controlled by the  magnitude of $q$,
%
 where a small  value of $q$ implies a small deviation from 
the collinear  ferromagnetic reference  state. 
Thus, the change $\Delta G(\vec{r},\vec{r}\,',E)$ in the Green function
can be obtained by solving the corresponding 
Dyson equation for $ G(\vec{r},\vec{r}\,',E)$ in linear approximation:
%
\begin{eqnarray}
 \Delta G(\vec{r},\vec{r}\,',E)
 & = &
\int_\Omega  d^3r'' \,G_0(\vec{r},\vec{r}\,'',E) \nonumber \\
&& \qquad  {\Delta V_{\vec{q}}(\vec{r}\,'')}\, G_0(\vec{r}\,'',\vec{r}\,',E) 
\;.
\label{Eq:linear-response}
\end{eqnarray}
%

Within the KKR-GF formalism
based on  multiple scattering theory the Green function  is represented in
real space by the scattering path operator 
${\underline{\tau}}^{n n'}$  together with the  regular
$Z^{n}_{\Lambda}(\vec{r},E)$ and irregular
$J^{n}_{\Lambda}(\vec{r},E)$ solutions of the single-site
Dirac equation (\ref{Eq:Dirac}) \cite{SPR-KKR6.3,EKM11,EBKM16}: 
\begin{widetext}
\begin{eqnarray}
G_0(\vec{r},\vec{r}\,',E) & = &
\sum_{\Lambda \Lambda'} 
Z^{n}_{\Lambda}(\vec{r},E)\,
                              {\tau}^{n n'}_{\Lambda\Lambda'}(E)\,
Z^{n' \times}_{\Lambda'}(\vec{r}\,',E)
 \nonumber \\
 & & 
-  \sum_{\Lambda} \Big[ 
Z^{n}_{\Lambda}(\vec{r},E)\, J^{n \times}_{\Lambda}(\vec{r}\,',E)\,
\Theta(r'-r)
+ J^{n}_{\Lambda}(\vec{r},E) \, Z^{n \times}_{\Lambda}(\vec{r}\,',E)\, \Theta(r-r')
\Big] \delta_{nn'} \; ,
\label{Eq:GF_KKR}
\end{eqnarray}
%
where the combined index $\Lambda =(\kappa,\mu)$ represents
the relativistic spin-orbit and magnetic quantum numbers
 $\kappa$ and $\mu$, respectively  \cite{Ros61}.

Inserting now  
Eq.\  (\ref{Eq:GF_KKR}) for the Green function $G_0(\vec{r},\vec{r}\,',E)$ 
and Eq.\  (\ref{Eq:perturb}) for the perturbation $\Delta V_{\vec{q}}(\vec{r})$ 
into  Eq.\  (\ref{Eq:linear-response}) for $\Delta G(\vec{r},\vec{r}\,',E)$
one can evaluate the linear  change in energy density  $K^{(1)}(\vec{r})$
from Eq.\  (\ref{K1_dens_new2}). Integrating over the whole space
one is led to the change in energy $K^{(1)}$: 
%
\begin{eqnarray} \label{Delta_K_Integral_1}
K^{(1)} &=& \int_\Omega d^3r \, K^{(1)}(\vec{r})  \nonumber  \\
&=&  - \sum_{ij} \frac{1}{\pi} \, \mbox{Im}\,\mbox{Tr} \int^\mu
dE \, (E - \mu)
    \int_{\Omega_i} d^3r_i \int_{\Omega_j} d^3r_j
  \,G_0(\vec{r}_j,\vec{r}_i,E) \,{\delta V(\vec{r}_i)}\,
    G_0(\vec{r}_i,\vec{r}_j,E)  \nonumber \\
&=& 
  - \sum_{ij} \frac{1}{\pi} \, \mbox{Im}\, 
\int^\mu dE  \, (E - \mu)  
\sum_{\Lambda_1\Lambda_2\Lambda_3\Lambda_4} {  O^j_{\Lambda_4\Lambda_1}(E)}  \,
        {\tau}^{j i}_{\Lambda_1\Lambda_2}(E) 
\Big[ { T^{i,x}_{\Lambda_2
          \Lambda_3}(E)} \, {\tau}^{i j}_{\Lambda_3\Lambda_4}(E)  \,
 \sin(\vec{q} \cdot (\vec{R}_i -  \vec{R}_j))
 \nonumber  \\
& &\qquad \qquad \qquad \qquad \qquad \qquad \qquad \qquad  \qquad \qquad \qquad 
+ { T^{i,z}_{\Lambda_2 \Lambda_3}(E)} \, {\tau}^{ij}_{\Lambda_3\Lambda_4}(E)\, 
(\cos(\vec{q} \cdot (\vec{R}_i -  \vec{R}_j)) - 1)  \Big] \;,
\label{Eq:K1-Integral}
\end{eqnarray}
%
where cell centered coordinates $\vec{r}_i =\vec{r} - \vec{R}_i$ 
have been used. 
The overlap integrals $O^{j}_{\Lambda\Lambda'}$ 
and matrix elements of the torque operator $T^{i,\alpha}_{\Lambda\Lambda'}$
occuring in Eq.\  (\ref{Eq:K1-Integral})
are defined as follows:\cite{EM09a}
%
\begin{eqnarray}
 O^{j}_{\Lambda\Lambda'} & = & \int_{\Omega_j} d^3r  \,
 Z^{j \times}_{\Lambda}(\vec{r},E) \, Z^{j}_{\Lambda'}(\vec{r},E)  \label{Eq:ME1}
  \\
 T^{i,\alpha}_{\Lambda\Lambda'} & = & \int_{\Omega_i} d^3r  \, Z^{i \times}_{\Lambda}(\vec{r},E)\, \Big[\beta \sigma_{\alpha} B_{xc}^i(\vec{r})\Big] \, Z^{i}_{\Lambda'}(\vec{r},E)\;.  \label{Eq:ME2}
\end{eqnarray}
%
Eq.\  (\ref{Eq:K1-Integral}) obviously  gives access  to the limit 
$ \lim_{q \to 0} \frac{\partial }{\partial  q_\alpha} K^{(1)}$
on the basis of KKR-GF based electronic structure calculations.
For the model Hamiltonian   Eq.\  (\ref{Eq:Heisenberg}),    on the other hand,
the corresponding quantity
 $\lim_{q \to 0} \frac{\partial }{\partial q_\alpha} E_{\rm DM}^{(1)}  $
is given by Eq.\  (\ref{Slope}).
Comparing both expressions and equating the corresponding terms
for each atom pair $(i,j)$, one arrives at the following
expression for the $y$ component of the DMI vector:
%
\begin{eqnarray}
  D^y_{ij} & = &
\left(-\frac{1}{2\pi} \right) \mbox{Im}\,\mbox{Tr} \int^\mu dE \,(E - \mu) 
       \bigg[ { \underline{O}^{j}(E)} 
     \,   \underline{\tau}^{j i}(E)  \, { \underline{T}^{i,x}(E)} \, \underline{\tau}^{i j}(E)  -  { \underline{O}^{i}(E)} \,  \underline{\tau}^{i
      j}(E) \, {\underline{T}^{j,x}(E)}  \,
        \underline{\tau}^{ji}(E)\bigg] \; , \label{Eq:DMI_Dij}
\end{eqnarray}
\end{widetext}
where the underline indicates matrices with respect to the spin-angular
index $\Lambda$. The scheme sketched here and called in the following
scheme III gives in a completely analogous way
the x-component of the DMI vector, $D^x_{ij}$. In this case one 
considers a spin spiral with the  spin moments
rotating  within the $y-z$ plane according to  
\begin{eqnarray}
\hat{m}_i =
\Big(0, \sin(\vec{q}\cdot\vec{R}_i),
\cos(\vec{q}\cdot\vec{R}_i)\Big)
\end{eqnarray}
and with the wave vector  $\vec{q} = (q,0,0)$ along the  axis  $\hat{x}$.
\medskip

Again, it should be noted that also in this case
 the  component $D^z_{ij}$ is undefined 
for the collinear  ferromagnetic reference  state with its
magnetic moments along $\hat{z}$, as it characterizes 
the interactions of the  components $m^x_{i(j)}$ and $m^y_{i(j)}$ 
of the local magnetic moments, which are equal to zero.
Accordingly, having the   magnetization oriented along  $\hat{x}$
one gets access to $D^z_{ij}$ and $D^y_{ij}$
while choosing the orientation along $\hat{y}$
one gets $D^z_{ij}$ and $D^x_{ij}$, respectively.
A more detailed comparison of the present approach with those reported
previously in the literature is given in Appendix A. 

\subsection{DMI vector within the micromagnetic approach}

For the sake of completeness, the micromagnetic
definition of the DMI based on the KKR formalism is considered next. The
DMI related energy is determined by the second term in Eq.\
(\ref{Eq:Micromanetic})  
\begin{eqnarray}
  E_{\rm DM} &=&  \sum_{\mu \nu} {D}^{\nu\mu} \bigg(\hat{m} \times  \frac{\partial
 \hat{m}}{\partial r_\mu} \bigg)_\nu \;.
\end{eqnarray}
A spin spiral described by the magnetization
\begin{equation*}
  \hat{m}(\vec{r}) =  \Big(\sin(\vec{q}\cdot\vec{r}),0, \cos(\vec{q}\cdot\vec{r})\Big)
\label{spiral_m}
\end {equation*}
%
results in an energy change due to the DMI according to
\begin{eqnarray}
  E^{(1)}_{\rm DM} &=&  \sum_{\mu} (\vec{D}^\mu \cdot \hat{y}) q_\mu =  \sum_{\mu} D^{y\mu} q_\mu
\end{eqnarray}
leading to the relation: 
\begin{equation}
 \frac{\partial  E^{(1)}}{\partial q_\alpha} \bigg\vert _{q \to 0} =
 \frac{\partial  E^{(1)}_{\rm DM}}{\partial q_\alpha} \bigg\vert _{q \to 0}
 = D^{y \alpha} \;.
\label{Eq:DMI_micromag}
\end {equation}

To derive an expression for this micromagnetic parameter one may start
again from the expression in Eq.\ (\ref{Delta_K_Integral_1}) rewritten
in the form: 
\begin{widetext}
\begin{eqnarray}
  K^{(1)} & = &-  \frac{1}{2\pi}\mbox{Im}\, \mbox{Tr}
  \sum_{ij} \int^\mu dE (E - \mu)  \nonumber  \\
    &&\times  \Bigg( \sin(\vec{q} \cdot (\vec{R}_i - \vec{R}_j)) \Big[\underbrace{ \underline{O}^{j}(E)\,
\underline{\tau}^{ji}(E)  \underline{T}^{i,x}(E)\, \underline{\tau}^{i
   j}(E) }_{K1} - \underbrace{ \underline{T}^{j,x}(E)\,
\underline{\tau}^{ji}(E)  \underline{O}^{i}(E)\, \underline{\tau}^{i
   j}(E) }_{K2} \Big]\nonumber \\
 &&+ [ \cos(\vec{q}
\cdot (\vec{R}_i - \vec{R}_j)) -1] \Big[\underbrace{  \underline{O}^{j}(E)\, \underline{\tau}^{ji}(E)\,
\underline{T}^{i,z}(E)\, \underline{\tau}^{ij}(E)  }_{K3} - \underbrace{  \underline{T}^{j,z}(E)\, \underline{\tau}^{ji}(E)\,
\underline{O}^{i}(E)\, \underline{\tau}^{ij}(E) }_{K4} \Big]   \Bigg) \;.
\label{Eq:Delta_K_Integral_mod}
\end{eqnarray}
To make connection with the micromagnetic approach it is
advantageous to use the representation of the scattering path operators in 
reciprocal space. Considering for the sake of simplicity one atom per
unit cell one has  $\underline{O}^{i}(E) = \underline{O}(E)$ and
$\underline{T}^{i,\alpha}(E) = \underline{T}^\alpha(E)$.
Calculating the  derivative  $\frac {\partial K^{(1)}}{\partial q_y}$
in the limit ${q \to 0}$, the first term $K1$ in
Eq.\  (\ref{Eq:Delta_K_Integral_mod}) yields
%
\begin{eqnarray}
K1 & \to & -\frac{1}{2\pi}\, \lim_{q \to 0} \frac {\partial }{\partial q_y}
  \Big[\mbox{Im}\, \mbox{Tr}
  \sum_{ij} \int^\mu dE \,(E - \mu)  \nonumber \\
 &&\times   \underline{O}(E)
 \frac {1}{\Omega_{BZ}}  \int d^3k \,
\underline{\tau}(\vec{k},E)\, e^{-i\vec{k}
\cdot \vec{R}_{ij}} \, \underline{T}^{z}(E) \frac {1}{\Omega_{BZ}}\int d^3k'\,
\underline{\tau}(\vec{k}',E)\, e^{i\vec{k}'\cdot \vec{R}_{ij}} \frac{1}{2i}\big(e^{i\vec{q}
\cdot \vec{R}_{ij}} - e^{-i\vec{q} \cdot \vec{R}_{ij}} \big)  \Big ] \nonumber \\
  &= &  -\frac{1}{\pi}\,  \lim_{q \to 0} \frac {\partial}{\partial q_y}
  \Big[\mbox{Im}\, \mbox{Tr}\,
   \frac{1}{2i}  \int^\mu dE \,(E - \mu)  \nonumber \\
  & & \times \Big\{ \underline{O}(E)\, 
 \frac {1}{\Omega_{BZ}}\int d^3k \,
\underline{\tau}(\vec{k},E)\,  \underline{T}^{z}(E)\,
\underline{\tau}(\vec{k} - \vec{q},E) - \underline{O}(E)\, \frac {1}{\Omega_{BZ}} \int d^3k \,
\underline{\tau}(\vec{k},E)\, \underline{T}^{z}(E)\,
 \underline{\tau}(\vec{k} + \vec{q},E)\Big\} \Big]  \nonumber \\
  &= &- \frac{1}{\pi} \mbox{Re}\, \mbox{Tr}\, \int^\mu dE \,(E - \mu) \frac {1}{\Omega_{BZ}} \int d^3k \, \underline{O}(E)\,
\underline{\tau}(\vec{k},E) \,
\underline{T}^{x}(E)\,\frac{\partial}{\partial k_y}\,
\underline{\tau}(\vec{k},E) \;. \nonumber  
\end{eqnarray}
%
In analogy, one gets for the second term $K2$ in
Eq.\  (\ref{Eq:Delta_K_Integral_mod}) the expression 
%
\begin{eqnarray}
K2 & \to &
- \frac{1}{\pi} \mbox{Re}\, \mbox{Tr}\, \int^\mu dE \,(E - \mu) \frac {1}{\Omega_{BZ}} \int d^3k \,  \underline{T}^{x}(E) \,
\underline{\tau}(\vec{k},E) \,
\underline{O}(E)\,\frac{\partial}{\partial k_y}\,
\underline{\tau}(\vec{k},E) \;. \nonumber  
\end{eqnarray}
%
Doing a corresponding transformation for the third term $K3$, one finds that
the derivative $\frac {\partial}{\partial q_y}$ vanishes in the limit  $q \to 0$: 
\begin{eqnarray}
K3 & \to & \frac{1}{\pi}\,  \lim_{q \to 0} \frac {\partial}{\partial q_y}
  \mbox{Im}\, \mbox{Tr}\,
   \int^\mu dE \,(E - \mu)  \nonumber \\
 &&\times \underline{O}(E)
 \Big[\frac {1}{\Omega_{BZ}}\int d^3k \,
\underline{\tau}(\vec{k},E) \, \underline{T}^{z}(E)\, \underline{\tau}(\vec{k} - \vec{q},E) +  \frac {1}{\Omega_{BZ}} \int d^3k \,
\underline{\tau}(\vec{k},E) \,\underline{T}^{z}(E)\,
 \underline{\tau}(\vec{k} + \vec{q},E) \Big] =  0\;. \nonumber 
\end{eqnarray}
%
Analogously, the term $K4$ vanishes as well.
With this, the $D^{yy}$ component of the DMI vector $\vec{D}^{y}$ is given
in the micromagnetic formulation by the expression:
\begin{eqnarray}
 D^{yy}  &=&  \lim_{q \to 0} \frac {\partial}{\partial q_y} K^{(1)} =
- \frac{1}{\pi} \mbox{Re}\, \mbox{Tr}\, \int^\mu dE \,(E - \mu)  \nonumber \\ 
 &&\times  \frac {1}{\Omega_{BZ}} \int d^3k \, \Big[ \underline{O}(E)\,
\underline{\tau}(\vec{k},E) \,
\underline{T}^{x}(E)\,\frac{\partial}{\partial k_y}\,
\underline{\tau}(\vec{k},E) -  \underline{T}^{x}(E) \,
\underline{\tau}(\vec{k},E) \,
\underline{O}(E)\,\frac{\partial}{\partial k_y}\,
\underline{\tau}(\vec{k},E)  \Big] \;.
\label{Eq:DYY_tau} 
\end{eqnarray}
\end{widetext}
The other DMI components  $D^{\alpha\beta}$ can be obtained in an
analogous way. 
This formulation gives access to a discussion
of the DMI in terms of specific features of the electronic band structure in a similar
way as suggested in Ref.\ \cite{KNA15}. On the other side, as this formulation is done within
the KKR-GF formalism, it allows to deal both with ordered and disordered
materials, where disorder may be treated using the coherent potential
approximation (CPA) alloy theory.

\section{Numerical results and discussion}

\begin{figure}[h]
\includegraphics[width=0.45\textwidth,angle=0,clip]{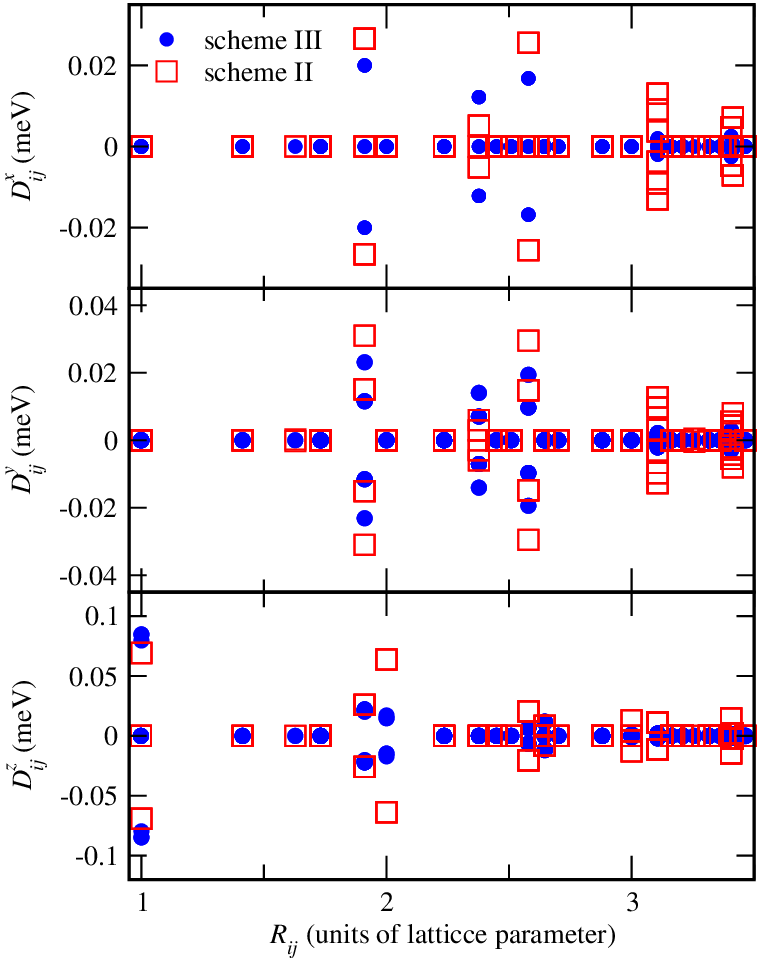}\;
\caption{\label{fig:CMP_DMI_Co} Components of the DMI vector $\vec{D}_{ij}$ for hcp
  Co. Results based on the present scheme (III) are compared with 
  results calculated using scheme II \cite{EM09a}. The 
calculations use a geometry with the direction of the DMI vector along the
magnetization direction in the latter case, and the DMI direction
perpendicular to the plane of the magnetization rotation in the former case. }   
\end{figure}

\begin{figure}[h]
\includegraphics[width=0.45\textwidth,angle=0,clip]{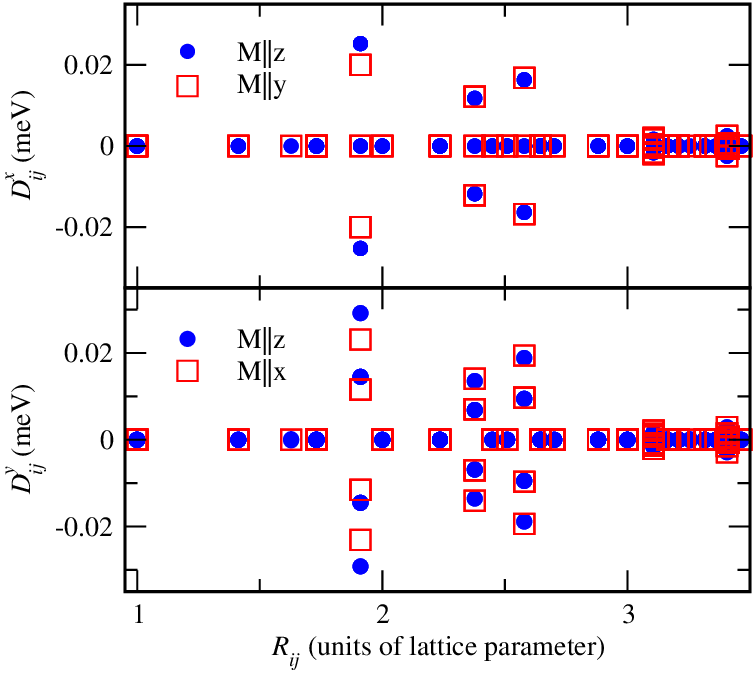}\;
\caption{\label{fig:DMI_Co_vs_M} $D^x_{ij}$ (top) and  $D^y_{ij}$ 
  (bottom) components of the DMI vector for hcp Co calculated via scheme
  III for the different reference states: for $\theta = 0$ and $\theta = \pi/2, \phi = 0$ in
  the case of  $D^x_{ij}$  (top) and  for $\theta = 0$ and $\theta = \pi/2, \phi = \pi/2$ in
  the case of  $D^y_{ij}$  (bottom). }   
\end{figure}

To illustrate the scheme introduced in section  \ref{THEO_II},
a comparison of the DMI
components calculated for hcp Co using the numerical schemes II
\cite{EM09a} and III (Eq.\ (\ref{Eq:DMI_Dij})) is shown in
Fig. \ref{fig:CMP_DMI_Co}. The symmetry properties of the system
allow non-zero interatomic DMI only within one sublattice. 
On the other hand, the elements of the micromagnetic tensor
$D^{\alpha\beta}$ calculated using Eq.\ (\ref{Eq:DYY_tau})  are zero as 
required for all systems exhibiting inversion symmetry.  
Note that different reference states are used
within these schemes. This is obviously one of the major sources 
leading to the observed deviation for the DMI parameters (see Appendix A
concerning this).
Another source responsible for the difference in the calculated
$\vec{D}_{ij}$ vectors is of course the different expression for the DMI
components in the present approach. 
Note also that disregarding the scheme used for the DMI calculations, 
a specific component can be determined for two different reference states
resulting in some difference between these values. This can be seen in
Fig. \ref{fig:DMI_Co_vs_M} representing the results for the $x$- and
$y$-components of the DMI vectors calculated for two different reference states 
for hcp Co within the scheme III discussed above.

\begin{figure}[h]
\includegraphics[width=0.45\textwidth,angle=0,clip]{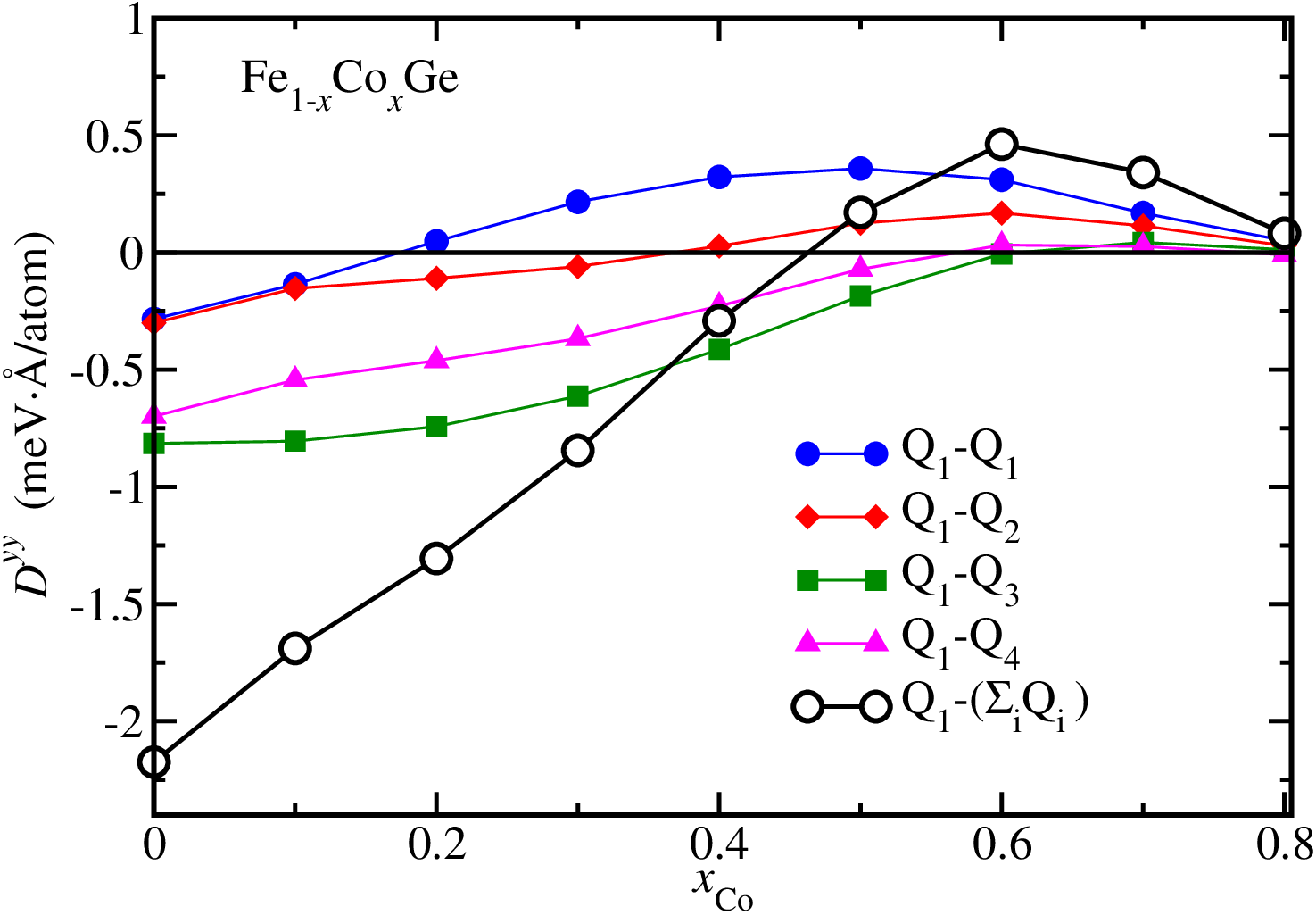}\;(a)
\includegraphics[width=0.45\textwidth,angle=0,clip]{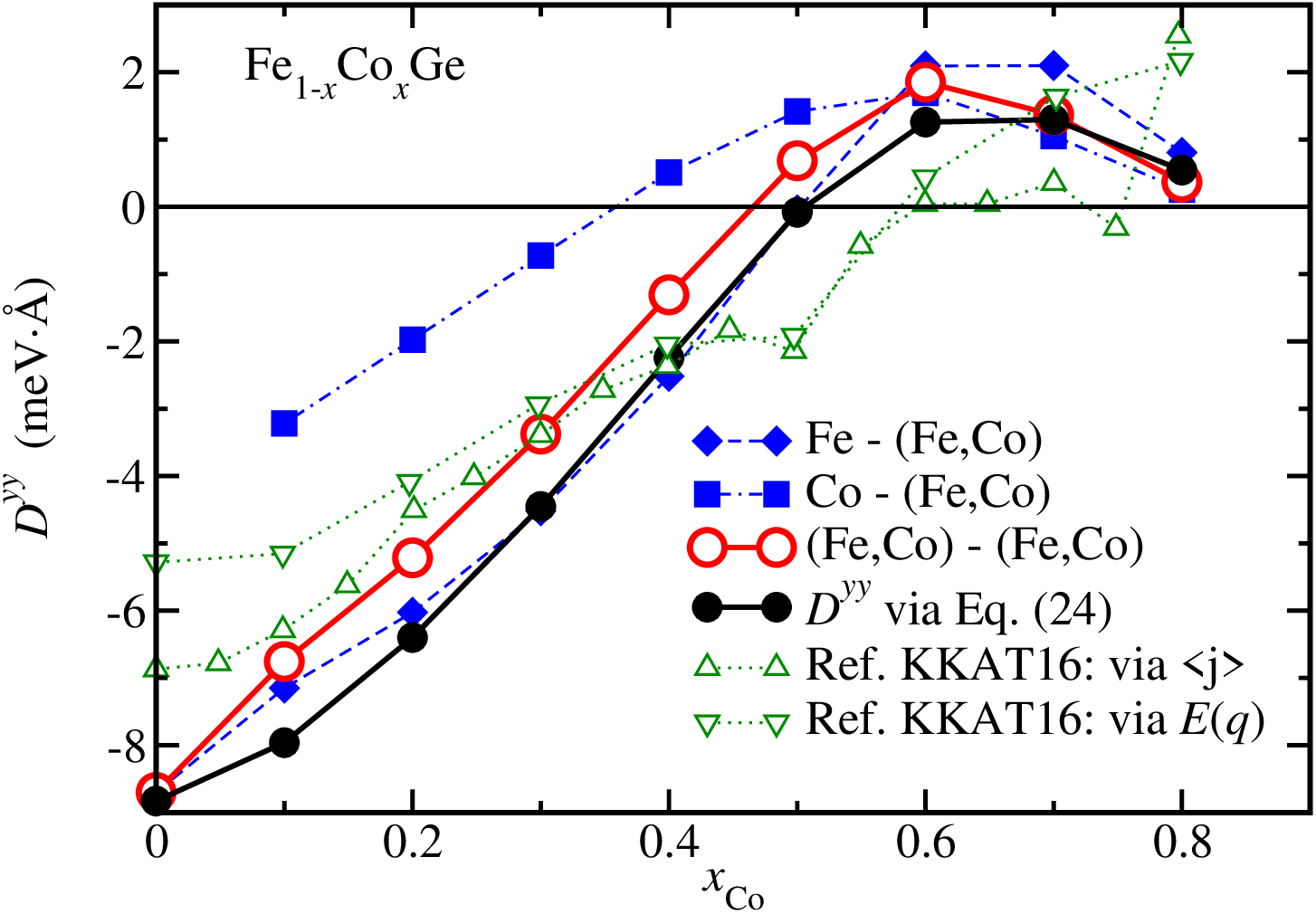}\;(b)
\caption{\label{fig:CMP_DMI_FeCoGe} The $D^{yy}$ component of the micromagnetic
  DMI vector as a function of Co concentration in  Fe$_{1-x}$Co$_{x}$Ge,
  representing the average value of interactions of Mn and Fe magnetic
  moments (represented per atom) with the magnetic moments of 4
  different (Fe,Co) sublattices (with $Q_m$ positions for (Fe,Co) atoms
  with $m = 1-4$) (a); and the average interactions of the magnetic moment on the (Fe,Co) site with the
 magnetic moment on all surrounding (Fe,Co) neighbor sites, represented
 per unit cell. The results are compared with the results of calculations
 based on Eq.\ (\ref{Eq:DYY_tau}) (solid circles)  as well as  with
 theoretical results of  
 Kikuchi et al. \cite{KKAT16}, shown by up- and down triangles, obtained
 in two different ways (for details see Ref. \cite{KKAT16})  (b).    
 }   
\end{figure}

As another example we consider the DMI in the substitutional 
alloy Fe$_{1-x}$Co$_{x}$Ge having the B20 crystal structure. 
Below the critical temperature $T_c$ and in the absence of an external magnetic
field this material shows a helimagnetic structure. 
The helix wave vector changes with the Co concentration 
reaching a minimum at $x \approx 0.6$, where the helix chirality 
changes sign \cite{GSA+14}. 
On the basis of electronic structure
calculations the $\vec{D}_{ij}$ interatomic interactions have been
calculated using Eq.\ (\ref{Eq:DMI_Dij}) up to $|\vec{R}_i - \vec{R}_j|
= R_{max} = 4a$ (with $a$ the lattice parameter). To treat the disorder on the
(Fe,Co) sublattice, the calculations have been performed using the
coherent potential approximation (CPA) alloy theory.
Because of the non-trivial crystal structure and chemical disorder in the
system, leading to a non-trivial analysis of the
concentration dependent behavior of the DMI,  
it is more convenient to use a micromagnetic description for the DMI.  
The values for the corresponding parameters can be evaluated using the interatomic $\vec{D}_{ij}$
interactions by comparing the derivatives of the energy in the atomistic
formulation, Eq.\ (\ref{Slope}), and in the micromagnetic formulation, Eq.\
(\ref{Eq:DMI_micromag}). This leads to the expression for the $y$-component of 
the micromagnetic DMI vector in terms of interatomic DMI vectors:
\begin{eqnarray}
 D^{yy}  &=&  \sum_{ij} D^y_{ij} \, (\vec{R}_j - \vec{R}_i)_y    \;.
\label{Eq:DMI_micmag_vs_Heis}
\end{eqnarray}
Figure \ref{fig:CMP_DMI_FeCoGe} represents the $D^{yy}$ element of the
micromagnetic DMI vector as a function of the Co concentration in
  Fe$_{1-x}$Co$_{x}$Ge. The values of the averaged interactions between the
  (Fe,Co) sites corresponding to different sublattices $m$,
  are shown in Fig. \ref{fig:CMP_DMI_FeCoGe}(a). 
As one can see, the various contributions $D^{yy}_{1-m}$ of the
sublattices $m$ to the total component $D^{yy} = \sum_{m} D^{yy}_{1-m}$
show a rather different concentration dependence. In particular, they
change sign at different Co concentration, while $D^{yy}$ changes
  sign at $x \approx  0.4$.  As can be seen in
  Fig. \ref{fig:CMP_DMI_FeCoGe}(b), the $D^{yy}$ component 
 of Co interaction with all surrounding (Fe,Co) sites changes sign at
a lower concentration, slightly above $x = 0.3$, while the corresponding
value of $D^{yy}$ for Fe becomes positive at $x \approx 0.5$.
The solid circles in Fig. \ref{fig:CMP_DMI_FeCoGe}(b) represent the
results of calculations based on  Eq.\ (\ref{Eq:DYY_tau}). As this
value accounts for all interactions in the system, it slightly
deviates from the  $D^{yy}$ value that accounts only for interactions within (Fe,Co)
sublattices, keeping however general trends concerning the concentration dependence.
 Despite certain differences, these results are in reasonable agreement with available
 experimental \cite{GSA+14} and theoretical \cite{KKAT16} data.

\section{Appendix A}

The expression for the DMI vector components derived within the present
work differs slightly from those given previously \cite{USPW03, EM09a}. 
This is because there is some freedom
concerning the scheme to map of the microscopic energy 
onto the extended Heisenberg Hamiltonian in Eq.\ (\ref{Eq:Heisenberg}).
This point is illustrated in the following.
Representing the components of the magnetic moments in spherical
coordinates, the DMI-part of the Heisenberg Hamiltonian has the form 
\begin{eqnarray}
  {\cal H}_{\rm DM}  &=& \sum_{ij} \vec{D}_{ij} \cdot (\hat{m}_i \times
  \hat{m}_j) 
  \nonumber  \\
  &=& \sum_{ij} {D}^x_{ij}\,
  [\sin(\phi_i)  - \sin(\phi_j)]\, \sin(\theta_i)\,\cos(\theta_j) 
  \nonumber  \\
  &&+ \sum_{ij} {D}^y_{ij}\,
  [\cos(\phi_j)  - \cos(\phi_i)]\, \sin(\theta_i)\,\cos(\theta_j) 
  \nonumber  \\
  &&+ \sum_{ij} {D}^z_{ij}\,
  [\cos(\phi_i) \,\sin(\phi_j) -
  \cos(\phi_j)\, \sin(\phi_i)]\, \nonumber \\
&&\times \sin(\theta_i)\,\sin(\theta_j) \;.
\label{Heisenberg_DM}
\end{eqnarray}
To simplify the discussion, let's consider a system with one sublattice
and focus on the last term with the
magnetization direction $\vec{m}||\hat{z}$ (i.e. $\theta = 0$ for all
local magnetic moments in the 
reference state). In this case the $D^z_{ij}$ component of the DMI vector can be 
represented by the second derivative of the energy with respect to the
angle $\theta_{i(j)}$:   
\begin{equation}
D^z_{ij}  = \frac{1}{2} \left[ \frac{\partial^2 E}{\partial \theta_i
    \partial \theta_j } \bigg|_{\substack{\phi_i = 0\\ \phi_j = \pi/2\\ \theta_{i(j)} = 0}}
  -  \frac{\partial^2 E}{\partial \theta_i \partial \theta_j }
  \bigg|_{\substack{\phi_j = 0\\ \phi_i = \pi/2\\ \theta_{i(j)} = 0}}
\right] \;.
\label{Eq:DM_def1}
\end {equation}
This means that the $D^z_{ij}$ component of the DMI vector is evaluated
as the energy difference between the reference state  
($\hat{m}_i||\hat{z}$ for all $i$) and the state with  
two magnetic moments having small components in the $x-y$ plane, which
are orthogonal to each other, giving a maximal energy change due to the DM 
interaction between these in-plane components.  

Similar to Eq.\  (\ref{K1_dens_new1}) an evaluation can also be done
considering the energy change due to tilting of two magnetic
moments. In this case one has:
\begin{widetext} 
\begin{eqnarray}
K_{ij}^{(1)}(\vec{r}) &=& -\frac{1}{\pi} \mbox{Im}\,\mbox{Tr}
\int^\mu dE\, [({\cal H} - \mu)\,  
  G(\vec{r},\vec{r}\,',E) - ({\cal H}_0 - \mu)\; G_0(\vec{r},\vec{r}\,',E) ]  \nonumber \\
&=&  -\frac{1}{\pi} \mbox{Im}\,\mbox{Tr} \int^\mu dE\, (H_0 - \mu)\;
\Delta G(\vec{r},\vec{r}\,',E) -\frac{1}{\pi} \mbox{Im}\,\mbox{Tr} \int^\mu dE\, (\delta V_i + \delta V_j)\,
 G_0(\vec{r},\vec{r}\,',E) \nonumber \\
&&-   \frac{1}{\pi}
\mbox{Im}\,\mbox{Tr}\int^\mu dE\,\int_\Omega d^3r''  \, 
 (\delta V_i(\vec{r}) + \delta V_j(\vec{r}))\,
G_0(\vec{r},\vec{r}\,'',E)\, (\delta V_i(\vec{r}\,'') +
\delta V_j(\vec{r}\,''))\, G_0(\vec{r}\,'',\vec{r}\,',E)  \;,
\label{Eq:K1_dens_old} 
\end{eqnarray}
\end{widetext}
with $\delta V_i  = \beta \vec{\sigma} \cdot (\hat{m}_i - \hat{z})
B_{xc}^i(\vec{r})$.
Eq.\  (\ref{Eq:K1_dens_old}) can be used for the evaluation of the
elements of the exchange coupling tensor by calculating the second energy
derivative $\frac{\partial^2 K^{(1)}}{\partial \alpha_1 \partial \beta_2}$
with $K^{(1)} = \int_{\Omega} K^{(1)}(r)d^3r$.
The first two terms yield 0, while the third term gives
the elements of the exchange coupling tensor similar to the one used in
\cite{EM09a}. For the particular case this leads to the $z$ component of
the DMI vector given by Eq.\  (\ref{Eq:DM_def1}), as provided by scheme
II \cite{EM09a}.  

To discuss an alternative possibility, let's represent the last term of 
Eq.\  (\ref{Heisenberg_DM}) as follows
\begin{eqnarray}
\sum_{ij} {D}^z_{ij}\, 
 \sin(\phi_j - \phi_i)\, \sin(\theta_i)\,\sin(\theta_j) \;. \nonumber 
\label{Heisenberg_DM}
\end{eqnarray}
In this case, taking the direction of magnetization within the plane, 
implying $\theta = \pi/2$, $D^z_{ij}$ gives the energy variation
due to tilting of two magnetic moments away from a collinear orientation by the 
small angles $\phi_j$ and $\phi_i$. In this case $D^z_{ij}$ can be
represented through the first derivatives of the energy with respect to
$\phi_j$ and $\phi_i$
\begin{eqnarray}
D^z_{ij}  &=& \frac{1}{2} \left[ \frac{\partial E}{\partial \phi_j}
  \bigg|_{\substack{\theta_i,\theta_j = \pi/2\\ \phi_i,\phi_j = 0 \\
    \hat{n} = \hat{z}}} -
  \frac{\partial E}{\partial \phi_i} \bigg|_{\substack{\theta_i,\theta_j = \pi/2\\ \phi_i,\phi_j = 0 \\
    \hat{n} = \hat{z}}} \right] \nonumber \\
 &=& -\frac{1}{2} \left[ R_{ji}^{xy} \bigg|_{\substack{\theta_{i(j)} =
      \pi/2\\ \phi_i,\phi_j = 0}} -  R_{ji}^{yx}
  \bigg|_{\substack{\theta_{i(j)} = \pi/2\\ \phi_i,\phi_j = 0}}
  \right] \;,
\label{Eq:DM_def2_1}
\end {eqnarray}
%
where $\hat{n}$ is the direction of the torque $ \vec{R}^{ij}_j$, and
$R^{z,ij}_i$ is the projection of the torque acting on magnetic moment
$\vec{m}_i$, originated due to the DMI and given by the expression
%
\begin{eqnarray}
 R_{ji}^{xy} &=&  \frac{1}{\pi}
\mbox{Im}\,\mbox{Tr}\int^\mu dE\, \nonumber \\
&&\times \underline{T}^{j,x}(E)  \,
\underline{\tau}^{ji}(E)\, \underline{T}^{i,y}(E)\,
 \underline{\tau}^{ij}(E) \Big] \;.
\label{K1_dens_old_torquez}
\end{eqnarray}
%
Here it is assumed that the magnetic moment is oriented along the $x$
direction, and $\underline{T}^{i,\alpha}(E)$ represents the matrix elements  of the operator
${\cal T}^{i,\alpha} = \beta \sigma_\alpha B_{xc}^i(\vec{r})$.



This alternative formulation is similar to the one worked out for the definition of the DMI by
Katsnelson et al. \cite{KKML10} and is in line with the derivation represented
above in this work Eq.\ (\ref{Eq:DMI_Dij}).

\section{Acknowledgement}

Financial support by the DFG via SFB 689 (Spinph\"anomene in reduzierten
Dimensionen) is gratefully acknowledged.


\begin{thebibliography}{21}
\expandafter\ifx\csname natexlab\endcsname\relax\def\natexlab#1{#1}\fi
\expandafter\ifx\csname bibnamefont\endcsname\relax
  \def\bibnamefont#1{#1}\fi
\expandafter\ifx\csname bibfnamefont\endcsname\relax
  \def\bibfnamefont#1{#1}\fi
\expandafter\ifx\csname citenamefont\endcsname\relax
  \def\citenamefont#1{#1}\fi
\expandafter\ifx\csname url\endcsname\relax
  \def\url#1{\texttt{#1}}\fi
\expandafter\ifx\csname urlprefix\endcsname\relax\def\urlprefix{URL }\fi
\providecommand{\bibinfo}[2]{#2}
\providecommand{\eprint}[2][]{\url{#2}}

\bibitem[{\citenamefont{Bogdanov and Hubert}(1994)}]{BH94}
\bibinfo{author}{\bibfnamefont{A.}~\bibnamefont{Bogdanov}} \bibnamefont{and}
  \bibinfo{author}{\bibfnamefont{A.}~\bibnamefont{Hubert}},
  \bibinfo{journal}{J. Magn. Magn. Materials} \textbf{\bibinfo{volume}{138}},
  \bibinfo{pages}{255 } (\bibinfo{year}{1994}), ISSN \bibinfo{issn}{0304-8853},
  \urlprefix\url{http://www.sciencedirect.com/science/article/pii/0304885394900469}.

\bibitem[{\citenamefont{Roszler et~al.}(2006)\citenamefont{Roszler, Bogdanov,
  and Pfleiderer}}]{RBP06}
\bibinfo{author}{\bibfnamefont{U.~K.} \bibnamefont{Roszler}},
  \bibinfo{author}{\bibfnamefont{A.~N.} \bibnamefont{Bogdanov}},
  \bibnamefont{and}
  \bibinfo{author}{\bibfnamefont{C.}~\bibnamefont{Pfleiderer}},
  \bibinfo{journal}{Nature} \textbf{\bibinfo{volume}{442}},
  \bibinfo{pages}{797} (\bibinfo{year}{2006}), ISSN \bibinfo{issn}{0028-0836},
  \urlprefix\url{http://dx.doi.org/10.1038/nature05056}.

\bibitem[{\citenamefont{Schulz et~al.}(2012)\citenamefont{Schulz, Ritz, Bauer,
  Halder, Wagner, Franz, Pfleiderer, Everschor, Garst, and Rosch}}]{SRB+12}
\bibinfo{author}{\bibfnamefont{T.}~\bibnamefont{Schulz}},
  \bibinfo{author}{\bibfnamefont{R.}~\bibnamefont{Ritz}},
  \bibinfo{author}{\bibfnamefont{A.}~\bibnamefont{Bauer}},
  \bibinfo{author}{\bibfnamefont{M.}~\bibnamefont{Halder}},
  \bibinfo{author}{\bibfnamefont{M.}~\bibnamefont{Wagner}},
  \bibinfo{author}{\bibfnamefont{C.}~\bibnamefont{Franz}},
  \bibinfo{author}{\bibfnamefont{C.}~\bibnamefont{Pfleiderer}},
  \bibinfo{author}{\bibfnamefont{K.}~\bibnamefont{Everschor}},
  \bibinfo{author}{\bibfnamefont{M.}~\bibnamefont{Garst}}, \bibnamefont{and}
  \bibinfo{author}{\bibfnamefont{A.}~\bibnamefont{Rosch}},
  \bibinfo{journal}{Nature Physics} \textbf{\bibinfo{volume}{8}},
  \bibinfo{pages}{301} (\bibinfo{year}{2012}),
  \urlprefix\url{http://dx.doi.org/10.1038/nphys2231}.

\bibitem[{\citenamefont{Neubauer et~al.}(2009)\citenamefont{Neubauer,
  Pfleiderer, Binz, Rosch, Ritz, Niklowitz, and B\"oni}}]{NPB+09}
\bibinfo{author}{\bibfnamefont{A.}~\bibnamefont{Neubauer}},
  \bibinfo{author}{\bibfnamefont{C.}~\bibnamefont{Pfleiderer}},
  \bibinfo{author}{\bibfnamefont{B.}~\bibnamefont{Binz}},
  \bibinfo{author}{\bibfnamefont{A.}~\bibnamefont{Rosch}},
  \bibinfo{author}{\bibfnamefont{R.}~\bibnamefont{Ritz}},
  \bibinfo{author}{\bibfnamefont{P.~G.} \bibnamefont{Niklowitz}},
  \bibnamefont{and} \bibinfo{author}{\bibfnamefont{P.}~\bibnamefont{B\"oni}},
  \bibinfo{journal}{Phys. Rev. Lett.} \textbf{\bibinfo{volume}{102}},
  \bibinfo{pages}{186602} (\bibinfo{year}{2009}),
  \urlprefix\url{http://link.aps.org/doi/10.1103/PhysRevLett.102.186602}.

\bibitem[{\citenamefont{Togawa et~al.}(2012)\citenamefont{Togawa, Koyama,
  Takayanagi, Mori, Kousaka, Akimitsu, Nishihara, Inoue, Ovchinnikov, and
  Kishine}}]{TKT+12}
\bibinfo{author}{\bibfnamefont{Y.}~\bibnamefont{Togawa}},
  \bibinfo{author}{\bibfnamefont{T.}~\bibnamefont{Koyama}},
  \bibinfo{author}{\bibfnamefont{K.}~\bibnamefont{Takayanagi}},
  \bibinfo{author}{\bibfnamefont{S.}~\bibnamefont{Mori}},
  \bibinfo{author}{\bibfnamefont{Y.}~\bibnamefont{Kousaka}},
  \bibinfo{author}{\bibfnamefont{J.}~\bibnamefont{Akimitsu}},
  \bibinfo{author}{\bibfnamefont{S.}~\bibnamefont{Nishihara}},
  \bibinfo{author}{\bibfnamefont{K.}~\bibnamefont{Inoue}},
  \bibinfo{author}{\bibfnamefont{A.~S.} \bibnamefont{Ovchinnikov}},
  \bibnamefont{and} \bibinfo{author}{\bibfnamefont{J.}~\bibnamefont{Kishine}},
  \bibinfo{journal}{Phys. Rev. Lett.} \textbf{\bibinfo{volume}{108}},
  \bibinfo{pages}{107202} (\bibinfo{year}{2012}),
  \urlprefix\url{http://link.aps.org/doi/10.1103/PhysRevLett.108.107202}.

\bibitem[{\citenamefont{Togawa et~al.}(2013)\citenamefont{Togawa, Kousaka,
  Nishihara, Inoue, Akimitsu, Ovchinnikov, and Kishine}}]{TKN+13}
\bibinfo{author}{\bibfnamefont{Y.}~\bibnamefont{Togawa}},
  \bibinfo{author}{\bibfnamefont{Y.}~\bibnamefont{Kousaka}},
  \bibinfo{author}{\bibfnamefont{S.}~\bibnamefont{Nishihara}},
  \bibinfo{author}{\bibfnamefont{K.}~\bibnamefont{Inoue}},
  \bibinfo{author}{\bibfnamefont{J.}~\bibnamefont{Akimitsu}},
  \bibinfo{author}{\bibfnamefont{A.~S.} \bibnamefont{Ovchinnikov}},
  \bibnamefont{and} \bibinfo{author}{\bibfnamefont{J.}~\bibnamefont{Kishine}},
  \bibinfo{journal}{Phys. Rev. Lett.} \textbf{\bibinfo{volume}{111}},
  \bibinfo{pages}{197204} (\bibinfo{year}{2013}),
  \urlprefix\url{http://link.aps.org/doi/10.1103/PhysRevLett.111.197204}.

\bibitem[{\citenamefont{Udvardi et~al.}(2003)\citenamefont{Udvardi, Szunyogh,
  Palot\'as, and Weinberger}}]{USPW03}
\bibinfo{author}{\bibfnamefont{L.}~\bibnamefont{Udvardi}},
  \bibinfo{author}{\bibfnamefont{L.}~\bibnamefont{Szunyogh}},
  \bibinfo{author}{\bibfnamefont{K.}~\bibnamefont{Palot\'as}},
  \bibnamefont{and}
  \bibinfo{author}{\bibfnamefont{P.}~\bibnamefont{Weinberger}},
  \bibinfo{journal}{Phys. Rev. B} \textbf{\bibinfo{volume}{68}},
  \bibinfo{pages}{104436} (\bibinfo{year}{2003}),
  \urlprefix\url{http://link.aps.org/doi/10.1103/PhysRevB.68.104436}.

\bibitem[{\citenamefont{Ebert and Mankovsky}(2009)}]{EM09a}
\bibinfo{author}{\bibfnamefont{H.}~\bibnamefont{Ebert}} \bibnamefont{and}
  \bibinfo{author}{\bibfnamefont{S.}~\bibnamefont{Mankovsky}},
  \bibinfo{journal}{Phys. Rev. B} \textbf{\bibinfo{volume}{79}},
  \bibinfo{pages}{045209} (\bibinfo{year}{2009}),
  \urlprefix\url{http://link.aps.org/doi/10.1103/PhysRevB.79.045209}.

\bibitem[{\citenamefont{Liechtenstein et~al.}(1987)\citenamefont{Liechtenstein,
  Katsnelson, Antropov, and Gubanov}}]{LKAG87}
\bibinfo{author}{\bibfnamefont{A.~I.} \bibnamefont{Liechtenstein}},
  \bibinfo{author}{\bibfnamefont{M.~I.} \bibnamefont{Katsnelson}},
  \bibinfo{author}{\bibfnamefont{V.~P.} \bibnamefont{Antropov}},
  \bibnamefont{and} \bibinfo{author}{\bibfnamefont{V.~A.}
  \bibnamefont{Gubanov}}, \bibinfo{journal}{J. Magn. Magn. Materials}
  \textbf{\bibinfo{volume}{67}}, \bibinfo{pages}{65} (\bibinfo{year}{1987}),
  \urlprefix\url{http://www.sciencedirect.com/science/article/pii/0304885387907219}.

\bibitem[{\citenamefont{Heide et~al.}(2008)\citenamefont{Heide, Bihlmayer, and
  Bl\"ugel}}]{HBB08}
\bibinfo{author}{\bibfnamefont{M.}~\bibnamefont{Heide}},
  \bibinfo{author}{\bibfnamefont{G.}~\bibnamefont{Bihlmayer}},
  \bibnamefont{and} \bibinfo{author}{\bibfnamefont{S.}~\bibnamefont{Bl\"ugel}},
  \bibinfo{journal}{Phys. Rev. B} \textbf{\bibinfo{volume}{78}},
  \bibinfo{pages}{140403} (\bibinfo{year}{2008}),
  \urlprefix\url{http://link.aps.org/doi/10.1103/PhysRevB.78.140403}.

\bibitem[{\citenamefont{Heide et~al.}(2009)\citenamefont{Heide, Bihlmayer, and
  Bl\"ugel}}]{HBB09}
\bibinfo{author}{\bibfnamefont{M.}~\bibnamefont{Heide}},
  \bibinfo{author}{\bibfnamefont{G.}~\bibnamefont{Bihlmayer}},
  \bibnamefont{and} \bibinfo{author}{\bibfnamefont{S.}~\bibnamefont{Bl\"ugel}},
  \bibinfo{journal}{Physica B: Condensed Matter}
  \textbf{\bibinfo{volume}{404}}, \bibinfo{pages}{2678 }
  (\bibinfo{year}{2009}), ISSN \bibinfo{issn}{0921-4526},
  \bibinfo{note}{proceedings of the Workshop: At the Frontiers of Condensed
  Matter IV. Current Trends and Novel Materials},
  \urlprefix\url{http://www.sciencedirect.com/science/article/pii/S0921452609004177}.

\bibitem[{\citenamefont{{Freimuth} et~al.}(2014)\citenamefont{{Freimuth},
  {Bl{\"u}gel}, and {Mokrousov}}}]{FBM14}
\bibinfo{author}{\bibfnamefont{F.}~\bibnamefont{{Freimuth}}},
  \bibinfo{author}{\bibfnamefont{S.}~\bibnamefont{{Bl{\"u}gel}}},
  \bibnamefont{and}
  \bibinfo{author}{\bibfnamefont{Y.}~\bibnamefont{{Mokrousov}}},
  \bibinfo{journal}{J. Phys.: Cond. Mat.} \textbf{\bibinfo{volume}{26}},
  \bibinfo{pages}{104202} (\bibinfo{year}{2014}), \eprint{1308.5983},
  \urlprefix\url{http://stacks.iop.org/0953-8984/26/i=10/a=104202}.

\bibitem[{\citenamefont{Koretsune et~al.}(2015)\citenamefont{Koretsune,
  Nagaosa, and Arita}}]{KNA15}
\bibinfo{author}{\bibfnamefont{T.}~\bibnamefont{Koretsune}},
  \bibinfo{author}{\bibfnamefont{N.}~\bibnamefont{Nagaosa}}, \bibnamefont{and}
  \bibinfo{author}{\bibfnamefont{R.}~\bibnamefont{Arita}},
  \bibinfo{journal}{Scientific Reports} \textbf{\bibinfo{volume}{5}},
  \bibinfo{pages}{13302} (\bibinfo{year}{2015}),
  \urlprefix\url{http://dx.doi.org/10.1038/srep13302}.

\bibitem[{\citenamefont{Engel and Dreizler}(2011)}]{ED11}
\bibinfo{author}{\bibfnamefont{E.}~\bibnamefont{Engel}} \bibnamefont{and}
  \bibinfo{author}{\bibfnamefont{R.~M.} \bibnamefont{Dreizler}},
  \emph{\bibinfo{title}{Density Functional Theory -- An advanced course}}
  (\bibinfo{publisher}{Springer}, \bibinfo{address}{Berlin},
  \bibinfo{year}{2011}),
  \urlprefix\url{http://www.springerlink.com/content/m246q2}.

\bibitem[{\citenamefont{Rose}(1961)}]{Ros61}
\bibinfo{author}{\bibfnamefont{M.~E.} \bibnamefont{Rose}},
  \emph{\bibinfo{title}{Relativistic Electron Theory}}
  (\bibinfo{publisher}{Wiley}, \bibinfo{address}{New York},
  \bibinfo{year}{1961}),
  \urlprefix\url{http://openlibrary.org/works/OL3517103W/Relativistic_electron_theory}.

\bibitem[{\citenamefont{{H.\ Ebert et al.}}(2012)}]{SPR-KKR6.3}
\bibinfo{author}{\bibnamefont{{H.\ Ebert et al.}}}, \bibinfo{howpublished}{{\em
  The Munich SPR-KKR package}, version 6.3,
  http://olymp.cup.uni-muenchen.de/ak/ebert/SPRKKR} (\bibinfo{year}{2012}),
  \urlprefix\url{http://olymp.cup.uni-muenchen.de/ak/ebert/SPRKKR}.

\bibitem[{\citenamefont{Ebert et~al.}(2011)\citenamefont{Ebert, K\"odderitzsch,
  and Min\'{a}r}}]{EKM11}
\bibinfo{author}{\bibfnamefont{H.}~\bibnamefont{Ebert}},
  \bibinfo{author}{\bibfnamefont{D.}~\bibnamefont{K\"odderitzsch}},
  \bibnamefont{and}
  \bibinfo{author}{\bibfnamefont{J.}~\bibnamefont{Min\'{a}r}},
  \bibinfo{journal}{Rep. Prog. Phys.} \textbf{\bibinfo{volume}{74}},
  \bibinfo{pages}{096501} (\bibinfo{year}{2011}),
  \urlprefix\url{http://stacks.iop.org/0034-4885/74/i=9/a=096501}.

\bibitem[{\citenamefont{Ebert et~al.}(2016)\citenamefont{Ebert, Braun,
  K\"odderitzsch, and Mankovsky}}]{EBKM16}
\bibinfo{author}{\bibfnamefont{H.}~\bibnamefont{Ebert}},
  \bibinfo{author}{\bibfnamefont{J.}~\bibnamefont{Braun}},
  \bibinfo{author}{\bibfnamefont{D.}~\bibnamefont{K\"odderitzsch}},
  \bibnamefont{and}
  \bibinfo{author}{\bibfnamefont{S.}~\bibnamefont{Mankovsky}},
  \bibinfo{journal}{Phys. Rev. B} \textbf{\bibinfo{volume}{93}},
  \bibinfo{pages}{075145} (\bibinfo{year}{2016}),
  \urlprefix\url{http://link.aps.org/doi/10.1103/PhysRevB.93.075145}.

\bibitem[{\citenamefont{Grigoriev et~al.}(2014)\citenamefont{Grigoriev,
  Siegfried, Altynbayev, Potapova, Dyadkin, Moskvin, Menzel, Heinemann, Axenov,
  Fomicheva et~al.}}]{GSA+14}
\bibinfo{author}{\bibfnamefont{S.~V.} \bibnamefont{Grigoriev}},
  \bibinfo{author}{\bibfnamefont{S.-A.} \bibnamefont{Siegfried}},
  \bibinfo{author}{\bibfnamefont{E.~V.} \bibnamefont{Altynbayev}},
  \bibinfo{author}{\bibfnamefont{N.~M.} \bibnamefont{Potapova}},
  \bibinfo{author}{\bibfnamefont{V.}~\bibnamefont{Dyadkin}},
  \bibinfo{author}{\bibfnamefont{E.~V.} \bibnamefont{Moskvin}},
  \bibinfo{author}{\bibfnamefont{D.}~\bibnamefont{Menzel}},
  \bibinfo{author}{\bibfnamefont{A.}~\bibnamefont{Heinemann}},
  \bibinfo{author}{\bibfnamefont{S.~N.} \bibnamefont{Axenov}},
  \bibinfo{author}{\bibfnamefont{L.~N.} \bibnamefont{Fomicheva}},
  \bibnamefont{et~al.}, \bibinfo{journal}{Phys. Rev. B}
  \textbf{\bibinfo{volume}{90}}, \bibinfo{pages}{174414}
  (\bibinfo{year}{2014}),
  \urlprefix\url{https://link.aps.org/doi/10.1103/PhysRevB.90.174414}.

\bibitem[{\citenamefont{Kikuchi et~al.}(2016)\citenamefont{Kikuchi, Koretsune,
  Arita, and Tatara}}]{KKAT16}
\bibinfo{author}{\bibfnamefont{T.}~\bibnamefont{Kikuchi}},
  \bibinfo{author}{\bibfnamefont{T.}~\bibnamefont{Koretsune}},
  \bibinfo{author}{\bibfnamefont{R.}~\bibnamefont{Arita}}, \bibnamefont{and}
  \bibinfo{author}{\bibfnamefont{G.}~\bibnamefont{Tatara}},
  \bibinfo{journal}{Phys. Rev. Lett.} \textbf{\bibinfo{volume}{116}},
  \bibinfo{pages}{247201} (\bibinfo{year}{2016}),
  \urlprefix\url{https://link.aps.org/doi/10.1103/PhysRevLett.116.247201}.

\bibitem[{\citenamefont{Katsnelson et~al.}(2010)\citenamefont{Katsnelson,
  Kvashnin, Mazurenko, and Lichtenstein}}]{KKML10}
\bibinfo{author}{\bibfnamefont{M.~I.} \bibnamefont{Katsnelson}},
  \bibinfo{author}{\bibfnamefont{Y.~O.} \bibnamefont{Kvashnin}},
  \bibinfo{author}{\bibfnamefont{V.~V.} \bibnamefont{Mazurenko}},
  \bibnamefont{and} \bibinfo{author}{\bibfnamefont{A.~I.}
  \bibnamefont{Lichtenstein}}, \bibinfo{journal}{Phys. Rev. B}
  \textbf{\bibinfo{volume}{82}}, \bibinfo{pages}{100403}
  (\bibinfo{year}{2010}),
  \urlprefix\url{https://link.aps.org/doi/10.1103/PhysRevB.82.100403}.

\end{thebibliography}

\end{document}